\documentclass[12pt]{article}
\catcode`\@=11

\@addtoreset{equation}{section}
\def\theequation{\arabic{equation}}
\def\theequation{\thesection\arabic{equation}}

\newcommand{\be}{\begin{equation}}
\newcommand{\ee}{\end{equation}}
\newcommand{\ba}{\begin{eqnarray}}
\newcommand{\ea}{\end{eqnarray}}

\def\part{\partial}

\def\@normalsize{\@setsize\normalsize{15pt}\xiipt\@xiipt
\abovedisplayskip 14pt plus3pt minus3pt%
\belowdisplayskip \abovedisplayskip
\abovedisplayshortskip  \z@ plus3pt%
\belowdisplayshortskip  7pt plus3.5pt minus0pt}
\def\small{\@setsize\small{13.6pt}\xipt\@xipt
\abovedisplayskip 13pt plus3pt minus3pt%
\belowdisplayskip \abovedisplayskip
\abovedisplayshortskip  \z@ plus3pt%
\belowdisplayshortskip  7pt plus3.5pt minus0pt
\def\@listi{\parsep 4.5pt plus 2pt minus 1pt
            \itemsep \parsep
            \topsep 9pt plus 3pt minus 3pt}}

\def\underline#1{\relax\ifmmode\@@underline#1\else
        $\@@underline{\hbox{#1}}$\relax\fi}
\@twosidetrue
\relax

\catcode`@=12

\catcode`\@=11

\def\section{\@startsection{section}{1}{\z@}{3.5ex plus 1ex minus
   .2ex}{2.3ex plus .2ex}{\large\bf}}
\def\thesection{\arabic{section}.}
\def\thesubsection{\arabic{section}.\arabic{subsection}}

\def\ps@headings{\def\@oddfoot{}\def\@evenfoot{}
\def\@oddhead{\hbox{}\hfill
        \makebox[.5\textwidth]{\raggedright\ignorespaces --\thepage{}--
        \hfill }}
\def\@evenhead{\@oddhead}
\def\subsectionmark##1{\markboth{##1}{}} }
\renewcommand{\subsection}[1]{\addtocounter{subsection}{1}
\vspace{2.5mm}\par\noindent {\em \thesubsection . #1}\par
 \vspace{0.5mm} }
\ps@headings

\catcode`\@=12

\relax

\def\figcap{\section*{Figure Captions\markboth
        {FIGURECAPTIONS}{FIGURECAPTIONS}}\list
        {Fig. \arabic{enumi}:\hfill}{\settowidth\labelwidth{Fig. 999:}
        \leftmargin\labelwidth
        \advance\leftmargin\labelsep\usecounter{enumi}}}
 \relax
\def\tablecap{\section*{Table Captions\markboth
        {TABLECAPTIONS}{TABLECAPTIONS}}\list
        {Table \arabic{enumi}:\hfill}{\settowidth\labelwidth{Table 999:}
        \leftmargin\labelwidth
        \advance\leftmargin\labelsep\usecounter{enumi}}}
 \relax
\def\reflist{\section*{References\markboth
        {REFLIST}{REFLIST}}\list
        {[\arabic{enumi}]\hfill}{\settowidth\labelwidth{[999]}
        \leftmargin\labelwidth
        \advance\leftmargin\labelsep\usecounter{enumi}}}
 \relax

\catcode`\@=11

\def\marginnote#1{}
\newcount\hour
\newcount\minute
\newtoks\amorpm
\hour=\time\divide\hour by60
\minute=\time{\multiply\hour by60 \global\advance\minute by-
\hour}
\edef\standardtime{{\ifnum\hour<12 \global\amorpm={am}%
    \else\global\amorpm={pm}\advance\hour by-12 \fi
    \ifnum\hour=0 \hour=12 \fi
    \number\hour:\ifnum\minute<100\fi\number\minute\the\amorpm}}
\edef\militarytime{\number\hour:\ifnum\minute<100\fi\number\minute}
\def\draftlabel#1{{\@bsphack\if@filesw {\let\thepage\relax
  \xdef\@gtempa{\write\@auxout{\string
    \newlabel{#1}{{\@currentlabel}{\thepage}}}}}\@gtempa
    \if@nobreak \ifvmode\nobreak\fi\fi\fi\@esphack}
     \gdef\@eqnlabel{#1}}
\def\@eqnlabel{}
\def\@vacuum{}
\def\draftmarginnote#1{\marginpar{\raggedright\scriptsize\tt#1}}
\def\draft{\oddsidemargin -.5truein
        \def\@oddfoot{\sl preliminary draft \hfil
        \rm\thepage\hfil\sl\today\quad\militarytime}
        \let\@evenfoot\@oddfoot \overfullrule 3pt
        \let\label=\draftlabel
        \let\marginnote=\draftmarginnote
   
\def\@eqnnum{(\theequation)\rlap{\kern\marginparsep\tt\@eqnlabel}%
\global\let\@eqnlabel\@vacuum}  }
\def\preprint{\twocolumn\sloppy\flushbottom\parindent 1em
        \leftmargini 2em\leftmarginv .5em\leftmarginvi .5em
        \oddsidemargin -.5in    \evensidemargin -.5in
        \columnsep 15mm \footheight 0pt
        \textwidth 250mmin      \topmargin  -.4in
        \headheight 12pt \topskip .4in
        \textheight 175mm
        \footskip 0pt
        
\def\@oddhead{\thepage\hfil\addtocounter{page}{1}\thepage}
        \let\@evenhead\@oddhead \def\@oddfoot{} \def\@evenfoot{}  }
\def\titlepage{\@restonecolfalse\if@twocolumn\@restonecoltrue\onecolumn
     \else \newpage \fi \thispagestyle{empty}\c@page\z@
        \def\thefootnote{\fnsymbol{footnote}} }
\def\endtitlepage{\if@restonecol\twocolumn \else  \fi
        \def\thefootnote{\arabic{footnote}}
        \setcounter{footnote}{0}}  
\catcode`@=12
\relax

\def\ps@headings{\def\@oddfoot{}\def\@evenfoot{}
\def\@oddhead{\hbox{}\hfill
        \makebox[.5\textwidth]{\raggedright\ignorespaces --\thepage{}--
        \hfill }}
\def\@evenhead{\@oddhead}
\def\subsectionmark##1{\markboth{##1}{}} }

\ps@headings

\relax

\def\firstpage#1#2#3#4#5#6{
\begin{document}


\begin{titlepage}
\nopagebreak
\title{\begin{flushright}
        \vspace*{-1.8in}
       {\normalsize ROM2F-2000/2}\\[-10mm]
        {\normalsize hep-th/0001077}\\[-4mm]
\end{flushright}
\vfill 
\vskip 4mm
{#3}}
\vskip -12mm
\author{ #4 \\[0.1cm] #5}
\maketitle
\vskip -9mm     
\nopagebreak 
\begin{abstract} {\noindent #6}
\end{abstract}
\vskip 36pt
\begin{center}
Based on talks presented at\vskip 12pt 
QFTHEP, Moscow, May 1999, at QG99, Villasimius, September 1999\\
Como 2001, Les Houches 2001, and at the 2001 Johns Hopkins Meeting
\end{center}
\begin{flushleft}
\rule{16.1cm}{0.2mm}
\today
\end{flushleft}
\thispagestyle{empty}
\end{titlepage}}

\date{}
\firstpage{3118}{IC/95/34} {\large\bf Open-string models with broken 
supersymmetry} 
{Augusto \  Sagnotti} 
{\small\sl Dipartimento di Fisica \\[-3mm]
\small\sl Universit\`a di Roma ``Tor Vergata''\\[-3mm]
\small\sl INFN, Sezione di Roma ``Tor Vergata''\\[-3mm]
\small\sl 
Via della Ricerca Scientifica 1\\[-3mm]
\small\sl I-00133 Roma \ ITALY} 
{I review the salient features of three classes of open-string models 
with broken supersymmetry. These suffice to exhibit, in relatively
simple settings, the two phenomena of ``brane supersymmetry''
and ``brane supersymmetry breaking''. 
In the first class of models, to lowest order 
supersymmetry is broken both in the closed and in the open 
sectors. In the second class of models, to lowest order supersymmetry is 
broken in the closed sector, but is {\it exact} in the open sector, at least
for the low-lying modes, and often for entire towers of string excitations.
Finally, in the third class of models, 
to lowest order supersymmetry is {\it exact} in the closed (bulk) sector, 
but is broken in the open sector. Brane supersymmetry 
breaking provides a natural solution to some old difficulties 
met in the construction of open-string vacua.
 }


\section{Broken supersymmetry and type-0 models}

In this talk I would like to review the key features of some open-string
models with broken supersymmetry constructed in \cite{bs,as,ads,ads2}. These 
models may be derived in a systematic
fashion from corresponding models of oriented closed strings \cite{cargese}, 
and once more display a surprising richness compared to them. Since the 
relevant techniques 
have been discussed at length in the original papers, I will not present
any explicit derivations. Rather, referring to some of the resulting vacuum 
amplitudes, I will try to illustrate how supersymmetry
can be broken at tree level in the bulk, on some branes or everywhere.

Closed-string models with broken
supersymmetry were among the first new examples considered in the last decade.
In particular, the type-0 models \cite{type0} provided the first non-trivial 
instances of modified GSO projections compatible with modular 
invariance. In order to describe their partition functions, I will begin 
by introducing some notation that will be used repeatedly in the following, 
defining the four level-one SO(8) characters
\ba
O_8 = { \vartheta_3^4 + \vartheta_4^4 \over 2 \eta^4} \quad , \quad
V_8 = { \vartheta_3^4 - \vartheta_4^4 \over 2 \eta^4} \quad , \nonumber \\
S_8 = { \vartheta_2^4 - \vartheta_1^4 \over 2 \eta^4} \quad , \quad
C_8 = { \vartheta_2^4 + \vartheta_1^4 \over 2 \eta^4} \quad , \label{ss2}
\ea
where the $\vartheta_i$ are Jacobi theta functions and $\eta$ is the
Dedekind function. In terms of these characters, and leaving aside the
contribution of the eight transverse bosonic coordinates, the type II models
are described by
\ba
{\cal T}_{IIA} &=& (V_8 - S_8)( \bar{V}_8 - \bar{C}_8) \quad , \\
{\cal T}_{IIB} &=& |V_8 - S_8|^2 \quad ,
\ea
while the type-0A and type-0B models are described by
\ba
{\cal T}_{0A} = |O_8|^2 + |V_8|^2 + S_8 \bar{C}_8 + C_8 \bar{S}_8 \quad , \\
{\cal T}_{0B} = |O_8|^2 + |V_8|^2 + |S_8|^2 + |C_8|^2 \quad .
\ea
In these expressions, the characters $(O_8,V_8,S_8,C_8)$ depend on 
$q = exp(i 2 \pi \tau)$, with $\tau$ the modulus of the torus, while their
conjugates depend of $\bar{q}$. All these characters have power series
expansions of the type
\begin{equation}
\chi(q) = q^{h - c/24} \ \sum_{n=0}^\infty \, d_n \, q^n \quad ,
\end{equation} 
where the $d_n$ are integers.
The low-lying spectra, essentially manifest in this notation, include 
in all cases the universal triple ($g_{\mu\nu}$, $B_{\mu\nu}$, $\phi$). 
The corresponding
states fill a generic transverse matrix, the direct product of 
the ground states of the
$V_8$ module and of its conjugate $\bar{V}_8$. In addition, the type-IIA
superstring 
has a Majorana gravitino and a Majorana spinor from the NS-R and
R-NS sectors and a vector and a three-form from the R-R sector.
The fermions result from pairs of Majorana-Weyl spinors of opposite
chiralities, ground states of $V_8 \bar{C}_8$ and $S_8 \bar{V}_8$, 
while the nature of the R-R bosons is determined by the direct product of the
two inequivalent spinor representations of SO(8). These are the 
ground states of $S_8$ and $\bar{C}_8$, and the product $8_s \times 8_c$
indeed decomposes into a vector and a three-form.
A similar reasoning shows that the type-IIB superstring has a complex
Majorana-Weyl gravitino and a complex Weyl fermion from the NS-R and
R-NS sectors, and a scalar, a two-form and a self-dual four-form from the
R-R sector. The spectra of the type-0 models are purely bosonic, and can be 
essentially deduced from these. Aside from the universal
triple ($g_{\mu\nu}$, $B_{\mu\nu}$, $\phi$), their low-lying excitations 
include a tachyon, the ground state of the $O_8 \bar{O}_8$ sector,
while their R-R sectors are two copies of the previous ones, and include
a pair of vectors and a pair of three-forms for the 0A model, and a pair
of scalars, a pair of two-forms and an unconstrained four-form for the 0B
model. Both type-0 models are clearly non-chiral, and are thus 
free of gravitational anomalies.

Let us now turn to the open descendants of the type-0 models. Their
structure is essentially determined by the Klein bottle
projection. Leaving aside the contributions of the transverse bosons,
the conventional choice, originally discussed in \cite{bs},
corresponds to
\ba
{\cal K}_{0A} &=& \frac{1}{2} ( O_8 + V_8 ) \quad , \\
{\cal K}_{0B} &=& \frac{1}{2} ( O_8 + V_8 - S_8 - C_8 ) \quad .
\ea
It eliminates the NS-NS two-form $B_{\mu\nu}$, but does not affect the 
tachyon. The effect of ${\cal K}$ can be simply summarized recalling that
a {\it positive} ({\it negative}) sign implies a {\it symmetrization} ({\it 
antisymmetrization})
of the sectors fixed under left-right interchange. This
unoriented projection is frequently called $\Omega$. For
instance, ${\cal K}_{0B}$ symmetrizes the two NS-NS sectors, described by
$|O_8|^2$ and $|V_8|^2$, and antisymmetrizes the two R-R sectors, eliminating
the NS-NS $B_{\mu\nu}$ and leaving only a pair of R-R two-forms. In addition,
the projected spectrum generally includes invariant combinations of all 
pairs of sectors interchanged by $\Omega$, that do not contribute 
to ${\cal K}$. Thus, the low-lying spectrum
of the projected 0A model includes also a R-R vector
and a R-R three-form. As usual, the characters in these direct-channel 
Klein-bottle amplitudes depend on $q \bar{q} = 
exp(- 4 \pi \tau_2 )$.

The open sector of the 0A model, described by \cite{bs}
\ba
{\cal A}_{0A} &=& \frac{n_B^2 + n_F^2}{2} (O_8 + V_8) \ - \ n_B n_F 
(S_8 + C_8)  \quad , \\
{\cal M}_{0A} &=& - \ \frac{n_B + n_F}{2} \hat{V}_8 \ - \ \frac{n_B - n_F}{2} 
\hat{O}_8 \quad ,
\ea
is not chiral and involves {\it two} different ``real'' charges, 
corresponding to orthogonal or symplectic groups. These enter the
partition functions via the dimensions, here $n_B$ and $n_F$, 
of the corresponding fundamental representations, and in general are
subject to linear relations originating from (massless) tadpole conditions. 
The low-lying modes are
simply identified from the contributions of ${\cal A}$ and ${\cal M}$.
For instance, the model contains two
sets of $n_B(n_B-1)/2$ and $n_F(n_F-1)/2$ vectors (corresponding to $V_8$), 
enough to fill the adjoint representations of a pair of orthogonal groups, 
tachyons (corresponding to $O_8$) in doubly (anti)symmetric representations
and fermions (corresponding to the R characters $S_8$ and $C_8$) in 
bi-fundamental representations. This description of open-string
spectra is also useful in Conformal Field Theory, where it provides a
convenient encoding of the spectrum of boundary operators in a generating
function of their multiplicities. In this case, if one insists on 
demanding the cancellation of all NS-NS tadpoles, the result is the
family of gauge groups SO$(n_B) \times$ SO$(n_F)$, with $n_B + n_F = 32$. 
Here ${\cal A}$  depends on $(q \bar{q})^{1/4}$ = 
$exp(-  \pi \tau_2 )$. On the other hand, ${\cal M}$  depends on 
$- (q \bar{q})^{1/4}$ = $exp(-  \pi \tau_2 + i \pi)$, but 
the ``hatted'' characters are redefined by suitable phases, and are thus real.

The open sector of the 0B model, described by \cite{bs}
\ba
{\cal A}_{0B} &=& \frac{n_o^2 + n_v^2 + n_s^2 + n_c^2}{2} V_8 \ + \ 
(n_o n_v + n_s n_c) O_8 \nonumber \\
&-& \ (n_v n_s + n_o n_c) S_8 \ - \ (n_v n_c + n_o n_s) C_8  \quad , \\
{\cal M}_{0B} &=& - \ \frac{n_o + n_v + n_s + n_c}{2} \hat{V}_8 \quad ,
\ea
involves four different ``real'' charges,  and
is {\it chiral} but free of anomalies, as a result of the R-R tadpole conditions
$n_o = n_v$ and $n_s = n_c$. All irreducible gauge and gravitational
anomalies cancel as a result of the R-R tadpole conditions, while the residual 
anomaly polynomial 
requires a generalized Green-Schwarz mechanism \cite{ggs,as}. If one 
insists on demanding the cancellation 
of all NS-NS tadpoles, not related to anomalies \cite{pc} as the previous ones,
one obtains the family of gauge groups 
SO$(n_o) \times$ SO$(n_v) \times $SO$(n_s)\times$ SO$(n_c)$, with 
$n_o + n_v + n_s + n_c = 64$. Despite their apparent 
complication, these open-string models are actually simpler than the 
previous ones, since the 0B torus amplitude corresponds to 
the ``charge-conjugation'' modular invariant. This circumstance 
implies a one-to-one correspondence between types of boundaries and types 
of bulk sectors typical of the ``Cardy case'' of boundary CFT \cite{cardy}. 
In equivalent terms, this model has four types of boundary states, 
in one-to-one correspondence with the chiral sectors of the bulk spectrum. 
The boundary
states of the 0A model are a bit subtler, since they are proper
combinations of these that do not couple to the R-R states, that
cannot flow in the transverse channel compatibly with 10D Lorentz 
invariance. Indeed, the product of the two
spinor representations $8_s$ and $8_c$ does not contain the identity,
and consequently a right-moving $S$ state cannot reflect into a left-moving 
$C$ state at a Lorentz-invariant boundary.

The modified Klein bottle projection
\be
{{\cal K}^\prime}_{0B} = \frac{1}{2} ( - O_8 + V_8 + S_8 - C_8 ) \quad ,
\ee
first proposed in \cite{as} as an amusing application of the results 
of \cite{pss}, removes the tachyon and the NS-NS two-form $B_{\mu\nu}$ 
from the closed spectrum, and leaves a {\it chiral} unoriented closed 
spectrum that comprises the $(g_{\mu\nu},\phi)$ 
NS-NS pair, together with a two-form, an additional scalar and a 
self-dual four-form
from the R-R sectors. The resulting open spectrum, described by
\ba
{\cal A}^\prime_{0B} &=& - \frac{n^2 + \bar{n}^2 + m^2 + \bar{m}^2}{2} C_8 + 
(n \bar{n} + m \bar{m}) V_8 \nonumber \\
&+& (n \bar{m} + m \bar{n}) O_8 - (m n + \bar{m} \bar{n}) S_8  \quad , \\
{\cal M}^\prime_{0B} &=&  \frac{m + \bar{m} - n - \bar{n}}{2} C_8 \quad , 
\ea
involves the ``complex'' charges of a pair of unitary groups, subject to
the R-R tadpole constraint $m - n = 32$, that eliminates all 
(non-Abelian) gauge and gravitational anomalies. The notation resorts to
pairs of multiplicities \cite{bs}, say $m$ and $\bar{m}$, to
emphasize the different roles of the fundamental and conjugate fundamental
representations of a unitary group U(m). The tadpole conditions
identify the numerical values of $m$ and $\bar{m}$, but once again one
can read the low-lying spectrum directly and conveniently from the amplitudes
written in this form. The choice $n=0$ selects
a U(32) gauge group, with a spectrum that is free of tachyons both in the
closed and in the open sectors. All irreducible (non-Abelian) gauge and 
gravitational anomalies cancel, while the residual anomaly polynomial 
requires a generalized Green-Schwarz mechanism \cite{ggs}. On the
other hand, the U(1) factor is anomalous, and is thus lifted by a 
ten-dimensional generalization of the
mechanism of \cite{w84}, so that the effective gauge group of this
model is SU(32). Here one does not have the option of eliminating all the
NS-NS tadpoles, and as a result a dilaton potential is generated.

These models have also been studied in some detail in \cite{bgab}, first with 
the aim of connecting them to the bosonic string, and more recently with 
the aim of relating them to
(non-supersymmetric) reductions of M theory. This last approach goes beyond
the perturbative analysis, and therefore has the potential of 
discriminating between the various options. According to \cite{bgab},
both types of $0B$ descendants admit a non-perturbative definition, while 
the $0A$ descendants do not. It would be interesting to take a closer look
at this relatively simple model and try to elicit some manifestation
of this phenomenon. 

Let us now spend a few words to summarize the key features of these 
descendants, where supersymmetry is
broken both in the closed and in the open sectors.
Whereas the first two models have tachyons both in the closed and in the open
sectors, the last results from a
non-tachyonic brane configuration of impressive simplicity. This feature
actually extends to lower-dimensional
compactifications, as first shown by Angelantonj \cite{carlo}. These type 0
models, and in particular the non-tachyonic one, have interesting 
applications \cite{adstype0} in the framework of the 
AdS/CFT correspondence \cite{ads/cft}. A simple generalisation of 
this setting allows one
to describe the branes allowed in these ten-dimensional models and in
their ``parent'' oriented closed models. These 
results, originally obtained by a number of authors, can be efficiently
described in this formalism as in \cite{dms}.

\section{Scherk-Schwarz deformations and brane supersymmetry}

We may now turn to the second class of models. These rest on
elegant extensions of the Kaluza-Klein reduction,
known as Scherk-Schwarz deformations \cite{ss}, that allow one to induce
the breaking of supersymmetry from the different behaviors of fermionic and
bosonic modes in the internal space. This setting, as adapted to the entire 
perturbative spectra of models of oriented closed strings in \cite{fkpz},
is the starting point for the constructions in \cite{bd,ads}.
I will confine my attention to particularly simple examples, related to
the reduction of the type IIB superstring on a circle of radius $R$
where the momenta or the windings are subjected
to 1/2-shifts, compatibly with modular invariance, in such a way that 
all massless fermions are lifted in mass. In these models, supersymmetry is
completely broken, but several more complicated open-string models, with 
partial breaking of supersymmetry are discussed in \cite{ads,bg}. 

This Scherk-Schwarz deformation generically introduces
tachyons, in the first case {\it (momentum shifts)} for $R < \sqrt{\alpha'}$, 
and in the second case {\it (winding shifts)} for
$R > \sqrt{\alpha'}$. The former choice \cite{bd,ads} is essentially a 
Scherk-Schwarz deformation of the low-energy field theory, here lifted to the 
entire string spectrum. On the other hand, the
latter \cite{ads} is a bit more subtle to interpret from a field theory perspective,
and indeed the resulting deformation of the spectrum is removed in the
limit of small radius $R$, that strictly speaking is inaccessible to
the field theory description.
Naively, in the
first case the open descendants should not present new subtleties. One 
would expect that the momentum deformations be
somehow inherited by the open spectrum, and this is indeed what happens.
On the other hand, naively the open spectrum should be insensitive to
winding deformations, simply because the available Neumann strings have 
only momentum excitations. Here
the detailed analysis settles the issue in an interesting way. The 
open spectrum is indeed affected, although in a rather subtle way, 
and supersymmetry is effectively
broken again, at the compactification scale $1/R$, but is {\it exact}
for the massless modes. 
In order to appreciate this result, let us present the closed-string
amplitudes for the two cases, here written in the Scherk-Schwarz basis,
\ba
{\cal T}_1 &=& Z_{m,2n} ( V_8 {\bar V}_8 + S_8 {\bar S}_8 ) +  
Z_{m,2n+1}( O_8
{\bar O}_8 + C_8 {\bar C}_8 )  \nonumber \\
&-& Z_{m+1/2,2n}( V_8 {\bar S}_8 + S_8 {\bar V}_8 ) -
Z_{m+1/2,2n+1}( O_8 {\bar C}_8 + C_8 {\bar O}_8 ) \label{ss6}
\ea
and
\ba
{\cal T}_2 &=& 
Z_{2m,n} ( V_8 {\bar V}_8 + S_8 {\bar S}_8 ) +  
Z_{2m+1,n}(O_8
{\bar O}_8 + C_8 {\bar C}_8 ) \nonumber \\
&-& Z_{2m,n+1/2}( V_8 {\bar S}_8 + S_8 {\bar V}_8 ) -
Z_{2m+1,n+1/2}( O_8 {\bar C}_8 + C_8 {\bar O}_8 ) \quad , \label{ss7}
\ea
and the corresponding Klein bottle projections
\ba
{\cal K}_1 &=& \frac{1}{2}  \ (V_8 - S_8) \ Z_m \ ,  \\
{\cal K}_2 &=& \frac{1}{2}  \ (V_8 - S_8) \ Z_{2m} + \frac{1}{2} \
(O_8 - C_8) \ Z_{2m+1} \quad . \label{klss}
\ea
In these expressions, $Z_{m,n}$ denotes the usual Narain lattice sum for
the circle
\be
Z_{m,n} = \sum_{m,n} \ 
q^{\frac{\alpha'}{4}(\frac{m}{R} + \frac{n R}{\alpha'})^2} \
\bar{q}^{\frac{\alpha'}{4}(\frac{m}{R} - \frac{n R}{\alpha'})^2} \quad ,
\ee
while, for instance, $Z_{2m,n}$ denotes the sum restricted to even momenta.
In a similar fashion, $Z_m$ in (\ref{klss}) denotes the restriction of the
sum to the momentum lattice.

In writing the corresponding open sectors, I will now eliminate 
several contributions,
restricting the charge configurations in such a way that no tachyons are introduced.
This is, to some extent, in the spirit of the previous discussion of the 10D 
U(32) model, but here one
can also cancel all NS-NS tadpoles. Moreover, I will take
into account the infrared subtlety discussed in \cite{pw}, 
that in the model with winding shifts leads the emergence of additional 
tadpoles in the singular limit $R \to 0$, where whole towers of
massive excitations collapse to zero mass. With this proviso, the
corresponding open spectra are described by
\ba
{\cal A}_1 &=& \frac{n_1^2 + n_2^2}{2} ( V_8 Z_m - S_8 Z_{m + 1/2} )
+ n_1 n_2 ( V_8 Z_{m + 1/2} - S_8 Z_m ) \quad ,  \nonumber \\ 
{\cal M}_1 &=& - \frac{ n_1 + n_2 }{2} ( {\hat
V}_8 Z_m - {\hat S}_8 Z_{m + 1/2} ) \quad , \label{ss8}
\ea
and
\ba
{\cal A}_2 &=& 
\frac{n_1^2 + n_2^2}{2} ( V_8 - S_8 ) Z_m
+ n_1 n_2 ( O_8 - C_8 ) Z_{m+1/2} \quad , \nonumber \\
{\cal M}_2 &=& - \frac{ n_1 + n_2 }{2} \, \left( {\hat V}_8 -
(-1)^m  {\hat S}_8 \right) Z_m \quad , 
\label{ss9}
\ea

As anticipated, the first model, with $n_1 + n_2 = 32$, is essentially a 
conventional Scherk-Schwarz deformation of the type-I superstring.
It can also describe the type I spectrum at a finite temperature 
related to the internal radius $R$. On the other hand, the second model,
where $n_1=n_2=16$, is more interesting, and displays the first novel 
phenomenon
reviewed here, ``brane supersymmetry'': although supersymmetry is broken
at the compactification scale by the Scherk-Schwarz deformation, 
the massless modes of the open sector fill complete supersymmetry multiplets.
I would like to stress that the breaking of supersymmetry in the massive 
spectrum of the second model can also be regarded as a deformation, now 
resulting from the unpairing of the Chan-Paton
representations for bosonic and fermionic modes with lattice 
excitations at alternate massive levels.
This is the simplest instance of the phenomenon that, following \cite{kt}, we
can call ``brane supersymmetry''.  Here the residual supersymmetry is
present only for the massless modes, but in more complicated models
it extends to entire sectors of
the open spectrum, as first shown in \cite{bg}.

T-dualities turn these descriptions into
equivalent ones, that can often have more intuitive appeal \cite{tdualop}. This is particularly
rewarding in the second case: a T-duality along the circle can turn the 
winding deformation into a momentum deformation orthogonal to the brane
responsible for the open-string excitations. It is then perhaps
simpler to accept the previous result that the deformation, now a
momentum shift orthogonal to the brane, does not affect its
massless excitations. A little more work \cite{ads} 
results in a duality argument
that associates the (now momentum) shift to the eleventh dimension of
M theory, thus realizing the proposal of \cite{dg}. Thus, 
as is often the case, a simple perturbative type I 
phenomenon has  a non-perturbative origin in the heterotic string
(and vice versa).

\section{Brane supersymmetry breaking}

I will now conclude by reviewing the third possibility afforded by these
constructions. Here I will follow \cite{ads}, concocting a six-dimensional
analogue of ``discrete torsion'' \cite{dt}. The construction of the 
resulting closed
string model is another application of the methods of \cite{pss},
in the same spirit as the construction of the 10D U(32) model. Starting
from the $T^4/Z_2$ U(16) $\times$ U(16) model \cite{bs2,gp}, one can 
revert the Klein-bottle projection for all twisted
states. This results in an unoriented closed spectrum with (1,0) supersymmetry,
whose massless excitations, aside from the gravitational multiplet,
comprise 17 tensor multiplets and 4 hypermultiplets. In \cite{ads2} it is
shown how this choice, described by
\ba
{\cal T} &=& \frac{1}{2} |Q_o + Q_v|^2 \Lambda + \frac{1}{2} |Q_o -
Q_v|^2 {\left|\frac{2 \eta}{\theta_2}\right|}^4  \\
&+& \frac{1}{2} |Q_s +
Q_c|^2 {\left|\frac{2 \eta}{\theta_4}\right|}^4
+ \frac{1}{2} |Q_s +
Q_c|^2 {\left|\frac{2 \eta}{\theta_3}\right|}^4 \quad , \nonumber \\
{\cal K} &=& \frac{1}{4} \left\{ ( Q_o + Q_v ) ( P + W ) -  2 \times
16 ( Q_s + Q_c ) \right\} \quad , \label{bsb2}
\ea
{\it does not} allow a consistent supersymmetric solution of the 
tadpole conditions. A consistent solution does exist \cite{ads2},
but requires the introduction of anti-branes, with the end
result that supersymmetry, exact to lowest order in the bulk, is necessarily
broken on their world volume. 
Hence the name ``brane supersymmetry breaking'' for
this peculiar phenomenon, that has the attractive feature of
confining the breaking of supersymmetry, and the resulting contributions
to the vacuum energy, to a brane, or to a collection of branes, that float in
a bath of supersymmetric gravity. In writing these expressions, I have 
introduced the $(1,0)$ supersymmetric characters \cite{bs}
\ba
Q_o &=& V_4 O_4 - C_4 C_4 \quad , \qquad Q_v = O_4 V_4 - S_4 S_4 \quad ,
\nonumber \\
Q_s &=& O_4 C_4 - S_4 O_4 \quad , \qquad Q_c = V_4 S_4 - C_4 V_4  \quad .
\label{bsb1}
\ea 
The two untwisted ones, $Q_o$ and $Q_v$, start with a vector multiplet and a
hypermultiplet, and are $Z_2$ orbifold breakings of $(V_8 - S_8)$. Out of
the two twisted ones $Q_s$ and $Q_c$, only $Q_s$ describes massless modes, 
that in this case
correspond to a half-hypermultiplet. The breaking of supersymmetry
is demanded by the consistency of String Theory. This can 
be seen rather neatly from the dependence of the transverse-channel
Klein bottle amplitude at the origin of the lattices on the sign $\epsilon$
associated to the twisted states
\be
\tilde{\cal K}_0 = \frac{2^5}{4} \biggl\{ Q_o \biggl( \sqrt{v}  \pm
\frac{1}{\sqrt{v}}\biggr)^2 + Q_v \biggl( \sqrt{v}  \mp
\frac{1}{\sqrt{v}}\biggr)^2 \biggr\} \ , \label{bsb3}
\ee
where the upper signs would correspond to the conventional case of 
the U(16) $\times$ U(16) model, while the lower signs correspond to the
model of eq. (\ref{bsb2}). Since the terms with different powers of
$\sqrt{v}$ are related by tadpole conditions to the multiplicities of the
$N$ and $D$ charge spaces, a naive solution of the model corresponding to
the upper signs would require a negative multiplicity, $D = - 32$.  

The open sector is described by
\ba
{\cal A} &=& \frac{1}{4} \biggl\{ (Q_o + Q_v) ( N^2 P  + D^2 W ) + 
2 N D (Q'_s + Q'_c) {\biggl(\frac{\eta}{\theta_4}\biggr)}^2  \label{a10}
\label{a13} \\
&+& (R_N^2 + R_D^2) (Q_o - Q_v) {\biggl(\frac{2
\eta}{\theta_2}\biggr)}^2 + 2 R_N R_D ( - O_4 S_4 - C_4 O_4 + V_4 C_4
+ S_4 V_4 ){\biggl(\frac{
\eta}{\theta_3}\biggr)}^2 \biggr\} \,  \nonumber \\
{\cal M} &=& - \frac{1}{4} \biggl\{ N P ( \hat{O}_4
\hat{V}_4  + \hat{V}_4 \hat{O}_4  - \hat{S}_4 \hat{S}_4
- \hat{C}_4 \hat{C}_4 ) -  D W ( \hat{O}_4
\hat{V}_4  + \hat{V}_4 \hat{O}_4  + \hat{S}_4 \hat{S}_4
+ \hat{C}_4 \hat{C}_4 ) \nonumber \\ &-&\!\!\!\!\! N( 
\hat{O}_4 \hat{V}_4 \!-\! \hat{V}_4 \hat{O}_4 \!-\! \hat{S}_4 \hat{S}_4
\!+\! \hat{C}_4 \hat{C}_4 )\left(
{2{\hat{\eta}}\over{\hat{\theta}}_2}\right)^2  \!\!+\! D( \hat{O}_4
\hat{V}_4 \!-\! \hat{V}_4 \hat{O}_4 \!+\! \hat{S}_4 \hat{S}_4
\!-\! \hat{C}_4 \hat{C}_4)\left(
{2{\hat{\eta}}\over{\hat{\theta}}_2}\right)^2  
\biggr\} \ . \nonumber
\ea
Supersymmetry is broken on the antibranes, and indeed the amplitudes 
involve new 
characters $Q'_s$ and $Q'_c$, that describe supermultiplets of a 
chirally flipped supercharge and may
be obtained from eq. (\ref{bsb1}) upon the interchange of $S_4$ and $C_4$,
as well as other non-supersymmetric combinations.
The tadpole conditions determine the gauge group 
$[$ SO(16) $\times$ SO(16) $]_9 \times  
[$ USp(16) $\times$ USp(16) $]_{\bar{5}}$, and the
$99$ spectrum is supersymmetric, with (1,0) vector
multiplets for the SO(16) $\times$ SO(16) gauge group and a
hypermultiplet in the ${\bf\!
(16,16,1,1)}$.
On the other hand, the ${\bar 5} {\bar 5}$ spectrum is
not supersymmetric and,
aside from the $[$ USp(16) $\times$ USp(16) $]$ gauge vectors, 
contains quartets
of scalars in the ${\bf (1,1,16,16)}$, right-handed Weyl 
fermions in the
${\bf (1,1,120,1)}$ and ${\bf (1,1,1,120)}$ 
and left-handed Weyl fermions in 
the ${\bf (1,1,16,16)}$.
Finally, the ND sector, also not supersymmetric, comprises doublets of
scalars in the ${\bf (16,1,1,16)}$ and in the 
${\bf (1,16,16,1)}$, and additional
(symplectic) Majorana-Weyl fermions in the ${\bf
(16,1,16,1)}$ and ${\bf (1,16,1,16)}$. These fields are a peculiar feature
of six-dimensional space time, where 
one can define Majorana-Weyl fermions, if the Majorana condition is 
supplemented by the conjugation in a pseudo-real representation.
All irreducible gauge and gravitational
anomalies cancel also in this model, while the residual 
anomaly polynomial 
requires a generalized Green-Schwarz mechanism \cite{ggs}.

The radiative corrections in this model are quite interesting, since they 
convey the (soft) breaking
to the gravitational sector. At any rate, the situation in which
a model requires the simultaneous presence of branes and
presents itself in at least two other
instances  \cite{massphd,ads2}, the four dimensional $Z_2 \times Z_2$ model 
with discrete torsion
and the four-dimensional $Z_4$ model. In both cases,
brane supersymmetry breaking allows
a solution of all tadpole conditions and a consistent definition
of the open descendants.

One can actually enrich these constructions, allowing for the simultaneous
presence of branes and antibranes \cite{sug,au,ads2}. These configurations are 
generically
unstable, and their instability reflects itself in the presence of
tachyonic excitations, a feature that we have already confronted in our
analysis of the ten-dimensional type-0 models. The internal lattice can
be used to lift in mass the tachyons, at least within certain ranges of
parameters for the internal geometry, that is actually partly {\it stabilized},
as a result of the different scaling behavior ($O(\sqrt{v})$ and
$O(1/\sqrt{v})$) of the contributions of
the different D$_p$ branes. Thus, for instance, 
starting from the type-IIB superstring, one can
introduce both branes and anti-branes at the price of having
tachyonic excitations in the open spectrum \cite{sug}. In addition, even with 
special
tachyon-free configurations, simply waiving the restriction to configurations
free of NS-NS tadpoles often gives new interesting models with broken
supersymmetry. The simplest setting is provided
again by the type-IIB superstring that, aside from the type-I superstring,
has an additional chiral tachyon-free descendant,
free of gauge and gravitational anomalies, but with broken supersymmetry 
and a USp(32) gauge group. In lower-dimensional models, more possibilities
are afforded by the internal lattice, that may be used to lift in mass some
tachyons, leading to stable vacuum configurations including both branes
and antibranes. Some of these \cite{aiq}, related to the 
four-dimensional $Z_3$ orientifold 
of \cite{abpss}, appear particularly interesting.

I would like to conclude by mentioning that brane
configurations similar to these have also been studied by several other
authors over the last couple of years, from a different vantage point,
following Sen \cite{sen}. Brane studies are reminiscent of
monopole studies in gauge theories of a classical Electrodynamics of charge
probes in a given external field, and are an interesting enterprise
in their own right. Open-string vacua are particular
brane configurations that are also vacuum configurations for a
perturbative construction, and thus
become exact solutions in the limit of vanishing string coupling. This,
in retrospect, makes them particularly attractive and instructive, and makes
their study particularly rewarding. In addition, they have very amusing
applications, in particular to issues related to the AdS/CFT correspondence, 
that we are only starting to appreciate.
\vskip 24pt
\begin{flushleft}
{\large \bf Acknowledgments}
\end{flushleft}

I am grateful to the Organizers for their kind invitations and to 
C. Angelantonj, I. Antoniadis, M. Bianchi, G. D' Appollonio, E. Dudas, 
G. Pradisi, Ya.S. Stanev for stimulating discussions and very enjoyable
collaborations on the topics reviewed in this talk, 
originally prepared for the QFTHEP Meeting, held in Moscow in June 1999
and for the QG99 Meeting, held in Villasimius in September 1999. With
slight corrections in the text and in the references, it is now
being contributed to the Proceedings of the 2001 Como Conference
on ``Statistical Field Theories'', of the 2001 Les Houches Summer Institute
on ``Gravity, gauge theory and strings'' and of the 2001 
Johns Hopkins Meeting, since my presentations
in all cases are within the spirit of this short review.
A forthcoming review article \cite{review} will provide a 
comprehensive derivation of these world-sheet constructions
more easily accessible to
interested readers. This work was partly supported by the EEC
contracts HPRN-CT-2000-001222, HPRN-CT-2001-00148 and by the INTAS contract
99-1-590.


\end{document}